\documentclass[%
  preprint,
 amsmath,amssymb,
 aps,
]{revtex4-1}

\usepackage{color}
\usepackage{graphicx}
\usepackage{dcolumn}
\usepackage{bm}
\newcounter{matriz}
\newenvironment{matriz}{\refstepcounter{matriz}\equation}{\tag{A\thematriz}\endequation}

\makeatletter 
\renewcommand{\thefigure}{S\@arabic\c@figure}
\makeatother

\begin{document}

\preprint{APS/123-QED}

\title{A stochastic model for gene transcription on \textit{Drosophila melanogaster} embryos}

\author{Guilherme N. Prata$^{\text{1}}$}
 \email{gnprata@cdcc.usp.br}

\author{Jos\'{e} Eduardo M. Hornos$^{\text{2}}$}%
 \email{Deceased}

\author{Alexandre F. Ramos$^{\text{1,3,4}}$}
 \email{alex.ramos@usp.br}
\affiliation{$^{\text{1}}$ Escola de Artes, Ci\^{e}ncias e Humanidades, Universidade de S\~{a}o Paulo, Avenida Arlindo B\'{e}ttio, 1000, Ermelino Matarazzo, S\~{a}o Paulo, SP, Brazil, CEP: 03828-000}%
\affiliation{$^{\text{2}}$ Instituto de F\'{i}sica de S\~{a}o Carlos, Universidade de S\~{a}o Paulo, Av. Trabalhador S\~{a}o-Carlense, 400, S\~{a}o Carlos, SP, Brazil, CEP: 13566-590}%
\affiliation{$^{\text{3}}$ Departamento de Radiologia -- Faculdade de Medicina, Universidade de S\~ao Paulo}
\affiliation{$^{\text{4}}$ N\'ucleo de Estudos Interdisciplinares em Sistemas Complexos, Universidade de S\~ao Paulo.}

\date{\today}

\begin{abstract}
  We examine immunostaining experimental data for the formation of the strip 2 of \textit{even-skipped} transcripts on \textit{D. melanogaster} embryos. An estimate of the factor converting immunofluorescence intensity units into molecular numbers is given. The analysis of the {\em eve} dynamics at the region of the stripe 2 suggests that the promoter site of the gene has two distinct regimes: an earlier phase when it is predominantly activated until a critical time when it becomes mainly repressed. That suggests proposing a stochastic binary model for gene transcription on \textit{D. melanogaster} embryos. Our model has two random variables: the transcripts number and the state of the source of mRNAs given as active or repressed. We are able to reproduce available experimental data for the average number of transcripts. An analysis of the random fluctuations on the number of {\em eve} and their consequences on the spatial precision of the stripe 2 is presented. We show that the position of the anterior/posterior borders fluctuate around their average position by $\sim 1 \%$ of the embryo length which is similar to what is found experimentally. The fitting of data by such a simple model suggests that it can be useful to understand the functions of randomness during developmental processes.

\begin{description}
\item[DOI]
\item[PACS numbers] 87.16.Yc, 87.10.Ed, 87.10.Mn
\end{description}
\end{abstract}

\maketitle


\section{\label{sec:level1}Introduction}

Gene regulation plays a key role on the formation of strikingly precise spatio-temporal patterns of expression of the genetic information stored on the DNA of the metazoans. Although the intracellular environment has a plethora of molecular species, the biochemical reactions are occurring with reactants present in low copy numbers. That leads to random fluctuations, as predicted by Delbr\"{u}ck \cite{Delbruck1940}, and it is intriguing that biological processes, such as development, are so reliable. One astonishing example is the \textit{D. melanogaster} segmentation process and the formation of gene products stripes, for example, the \textit{even-skipped} mRNA's (here on just \textit{eve}) stripe 2 \cite{Small1992}. The application of experimental techniques was successful on the identification of the reactants regulating the stripe 2 formation. However, a theoretical approach has played a key role on the determination of the interactions among regulatory proteins and specific DNA sites controlling \textit{eve}'s gene expression \cite{Janssens2006, Kim2013}. Similarly, it is fair to expect that the use of the appropriate theoretical machinery shall play a useful role on the investigation of the noise effects or noise suppression on developmental processes. 

Despite fluorescence techniques have been applied to the detection of the noise in prokaryotic \cite{Elowitz2000,Cai2006} or eukaryotic \cite{Blake2003} cells, the stochasticity in metazoans stands as a deeper challenge to experimentalists and to theoreticians. The stochastic modeling of biochemical reactions have been done using the Langevin technique \cite{Aurell2002}, exact stochastic simulations \cite{Gillespie1976} or master equations \cite{Kepler2001,Thattai2001,Sasai2003}. The Langevin technique is useful for the analysis of average and variance, but may fail when bi-modal distributions appear. Simulations based on the Gillespie's algorithm may generate the complete distributions with the requirement of strong computational power and a full characterization of the analyzed chemical reactions \cite{Arkin1998,Wu2007}. However, the intracellular biochemical mechanisms of metazoans are still under investigation. Hence, a coarse-grained approach based on simple analytically solvable models would be welcome. Here, master equations for small sets of effective chemical reactions may be proposed and have their exact solutions obtained. Although this phenomenological approach would not help on the search for the complete set of chemical reactions taking place inside the cell, it may be powerful on effectively establishing the outcomes of those reactions. Furthermore, exactly solvable models might be used as building blocks for the description of more complex phenomena, such as noise transmission on gene networks.

In this manuscript we propose a spatial stochastic binary model for gene transcription on \textit{fruit fly} embryos. We apply our model for the dynamics of the formation of the stripe 2 of {\em eve} and compare our predictions with available experimental data \cite{Janssens2006,Surkova2008a}. Since immunostaining is used for detection of mRNA's we propose a strategy to convert immunofluorescence into absolute numbers. The analysis of data around the peak of the stripe 2 suggests that the dynamics of the number of transcripts is due to an mRNA source operating in two states. Therefore, we construct a spatial stochastic binary model for the dynamics of the state of the mRNA source and for the mRNA numbers. The transcription dynamics occurs in two phases: the first has the mRNA's source predominantly active whereas it is mostly repressed through the second phase. We are able to reproduce the experimental data for the dynamics of the average number of mRNA's \cite{Janssens2006}. Furthermore, we calculate the variation on the position of the anterior and posterior borders of the stripe 2 due to the fluctuations on the amount of {\em eve} and it turns out that it is similar to the one observed experimentally \cite{Surkova2008a}. Those results suggest that our model can be a useful tool for investigation of noise and noise propagation on fruit flies gene networks.

\section{Immunofluorescence data analysis}

We start this section by presenting our estimation on the factor to convert the immunofluorescence intensities into mRNA numbers. The converted data is used to present mRNA profiles. The dynamics of the {\em eve} around the peak of expression of the stripe 2 is analyzed. The observed dynamics corresponds to a promoter site operating in the on and off states.

\subsection{Transcription data analysis}
\begin{figure*}
\centering
\includegraphics[width=\textwidth]{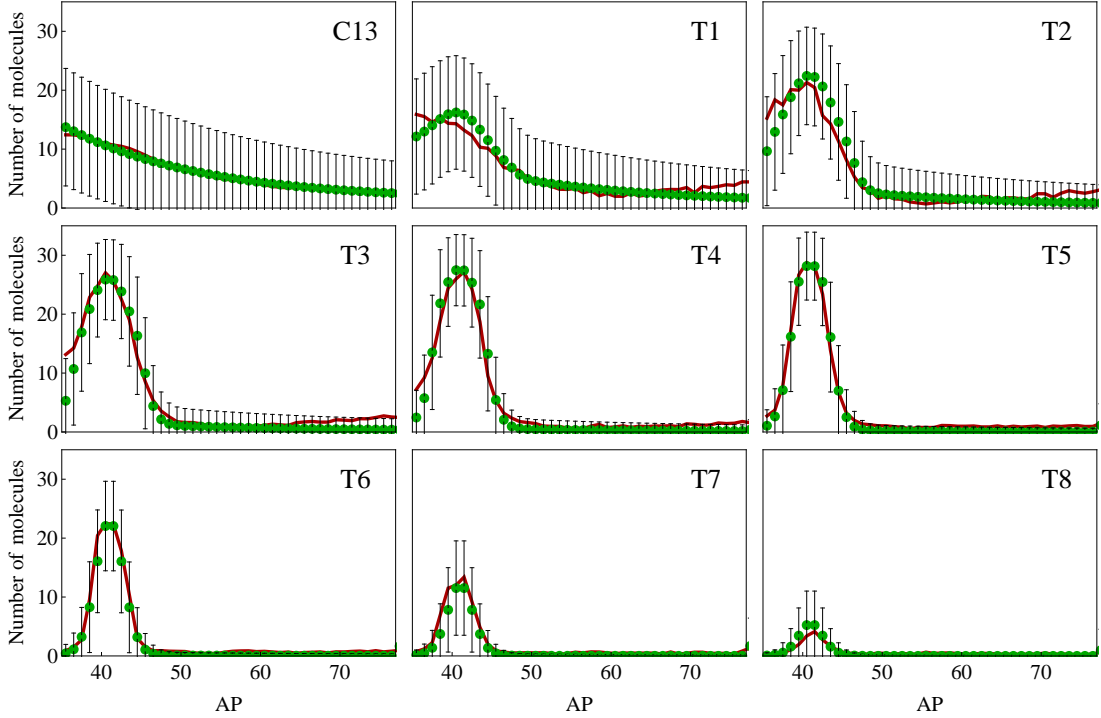}
\caption{\label{Fig1} 
The mRNA profiles of the {\em eve}'s stripe 2 is presented for C13 and for the eight time classes of C14, ending by the onset of gastrulation. The red lines indicate the experimental data while average mRNA numbers predicted by the stochastic binary model for gene transcription in fruit-flies is given by the solid green circles. The standard deviation on the mRNA numbers at each position along the AP axis is represented by the vertical black bars.
}
\end{figure*}

The experimental data analyzed in this manuscript has been presented in a recent paper \cite{Janssens2006}. Immunofluorescence intensities generated by the \textit{lacZ} transcription regulated by the enhancer controlling the eve gene of a \textit{D. melanogaster} embryo are provided. The immunofluorescence intensities refer to a strip along the AP axis of the embryo of width corresponding to 10\% of the dorsal-ventral embryo axis length. The intensity of immunofluorescence is indicated for a domain with length corresponding to one percent of the egg length along the AP axis. The expression of the genes involved in the \textit{Drosophila} embryo segmentation process can be considered uniform along the dorsal-ventral axis. Hence, the segmentation processes can be approximated as one dimensional along the AP axis. Finally, the investigation of the formation of \textit{eve}'s stripe 2 can be performed analyzing experimental data available for the positions from 35.5\% of the egg length (EL) to 76.5\% EL. The experimental data are grouped into nine different time classes, namely, C13 and T1, \dots, T8. {
  Each temporal class lasts 6.5 minutes (or 390 seconds) on average \cite{Janssens2006,Surkova2008a,Surkova2008b}.
}

In Ref. \cite{Janssens2006} the data for mRNA concentrations is available in immunofluorescence unities. To compare the data with our model predictions, we have converted immunofluorescence unities to absolute numbers. To the best of our knowledge, there is no information on the literature about the typical amount of \textit{eve} mRNA molecules around a nucleus during \textit{D. melanogaster} embryos development. Furthermore, we could not find conversion factors to transform immunofluorescence units into molecule numbers. Hence, we have estimated the number of \textit{eve} mRNA's considering information about the transcription of \textit{eve}, \textit{ftz}, and \textit{Ubx} genes,{
 all of them involved in the segmentation process of the {\em D. melanogaster} embryo during the blastoderm phase \cite{Poustelnikova2004,Pisarev2008,Surkova2008a,Surkova2008b,Myasnikova2005,Roark1985,Pareh2009,Liang1998,Muller1992,Tremml1992,Griffiths2008,Carroll1988,Martinez1985,Mahowald1985}.

The Ref. \cite{Janssens2006} shows the maximum fluorescence corresponding to the {\em eve} molecules at position 40.5\% EL at T5 instant. At this instant and position, the \textit{eve} mRNA number is assumed to be  $\sim$28.75. This quantity was computed considering the mRNA density to be 0.2 molecule/${\mu m}^3$ \cite{Pareh2009} and the volume of a nucleus during the cycle 14 as $\sim$143 ${\mu m}^3$ \cite{Bergman2008}. A linear extrapolation was carried out for the remaining fluorescence intensities along the embryo.   For our estimate, we assume what follows: (\textit{i}) \textit{eve} and \textit{ftz} mRNA concentrations are of same order of magnitude\cite{Poustelnikova2004,Pisarev2008,Surkova2008b,Myasnikova2005}; (\textit{ii}) abundances of \textit{ftz} and \textit{Ubx} mRNA’s are comparable\cite{Roark1985}; (\textit{iii}) the concentration of \textit{Ubx} mRNA molecules is 50 molecules/250 ${\mu m}^3$ at cycle 11\cite{Pareh2009}; (\textit{iv}) there is a linear relation between measured fluorescence intensity (without background signal) and the number of mRNA molecules. We adopt 50 molecules/250 ${\mu m}^3$ as reference value for the density of \textit{eve} mRNA molecules and the diameter of a nucleus of the \textit{Drosophila} embryo during cycle 14 to be 6.5 $\mu m$ \cite{Bergman2008}.
}

The Figure \ref{Fig1} shows a comparison between experimental data and theoretical predictions of the temporal evolution of the average number of {\em eve} from time class C13 to T8 along the AP axis. The experimental data is shown in red while the green lines are showing the model's predictions for the average mRNA numbers. The vertical bars are indicating the standard deviation on the mRNA numbers as predicted by our model. Each graph exhibits the amount of \textit{eve} transcripts at a given time class along the AP axis of the embryo. The horizontal axis of the plot gives the position of a nucleus as \% of the EL while the number of mRNA molecules is shown at the vertical axis.

\subsection{A two phases transcription dynamics}

\begin{figure*}
 \centering
 \includegraphics[scale=0.8]{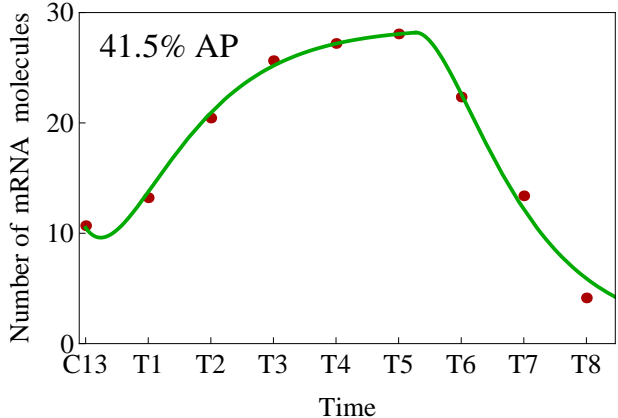}
 \caption{\label{Fig2} The two phases character of the dynamics of the {\em eve}'s gene transcription at the position 41.5 \% of EL is shown here. The experimental data is given by the solid red circles while the green line indicates the average mRNA number as predicted by our model. }
\end{figure*}

The Figure \ref{Fig2} shows the dynamics of the average number of \textit{eve} molecules at time instants from C13 to T8 at the position 41.5\% EL. The graph shows red solid circles indicating the experimental data and a green line indicating the average number of mRNA molecules as predicted by our model. The horizontal axis represents the time and the vertical axis shows the mRNA number.

Inspection is enough to verify that the amount of mRNA increases from C13 to T5. At T5 the amount of \textit{eve} mRNA's reaches a maximum and starts decreasing towards zero until T8. That dynamics suggests that the \textit{eve} gene is transcribed into two distinct phases: during the first phase the promoter site of the gene is predominantly in active state while it remains mostly repressed during the second phase. A similar time evolution pattern is observed along the AP axis of the embryo and can be considered as a signature of the formation of the \textit{eve} stripes. Therefore, we extend our conclusions for the position 41.5\% EL to the whole AP axis. We notice, however, that the time extent of both the increase and decrease phases vary along the AP axis.

\section{A coarse-grained approach for stochastic transcription}

  In this section we introduce a spatial stochastic binary model for the gene transcription on \textit{D. melanogaster} embryos. We apply the model for the expression of the {\em eve} gene by considering the formation of the stripe 2. The dynamics of the expression of the {\em eve} gene is a two phases process. The spatial profile of the critical times of the switch from the first phase to the second is presented. The fluctuations on the mRNA numbers and their implications on the variation of the position of the borders of the stripe 2 of {\em eve} are investigated.

\subsection{A spatial stochastic binary model for transcription in {\em Drosophila}}

For modeling purposes, we start considering a finite one-dimensional lattice with each node representing a cellular nucleus. The length of the lattice corresponds to the egg length (EL) at the AP direction. Each nucleus is labeled by a number $i$ which corresponds to its position at the AP axis in percentage of EL, namely, $i$ = 0.5, 1.5, \dots, 99.5. We assume the existence of a single copy of the gene per nucleus and we are interested on the number of mRNA's ($n_i$) present at the $i$-th nucleus. From here on we shall interpret the states of the mRNA source effectively as the promoter state, as its state regulates the transcription rate. We compare model predictions and experimental data assuming direct correspondence of spatial coordinates. In a given nucleus, the transcription may occur at the active (\textit{on}) and repressed (\textit{off}) gene states, respectively, with rates $k$ and $\chi k$ ($0\le \chi < 1$). The promoter transition to the active (or repressed) state at the $i$-th nucleus is indicated by the switching rates $f_i$ (or $h_i$).

We describe the dynamics of the system in terms of the probability, $\alpha_{n_i,i}(t)$ or $\beta_{n_i,i}(t)$, of finding $n_i$ mRNA's at the $i$-th nucleus and the promoter site being, respectively, active or repressed. The dynamics of these probabilities is controlled by the master equations, namely
\begin{eqnarray}
\label{mastereq}
\frac{d \alpha_{n_i,i}(t)}{dt} &=& k[\alpha_{n_i-1,i}(t)-\alpha_{n_i,i}(t)] 
                               + \rho[(n_i+1)\alpha_{n_i+1,i}(t)-n_i\alpha_{n_i,i}(t)] 
                               - h_i\alpha_{n_i,i}(t) + f_i\beta_{n_i,i}(t) \nonumber \\ \\\nonumber
\frac{d \beta_{n_i,i}(t)}{dt}  &=& k\chi[\beta_{n_i-1,i}(t)-\beta_{n_i,i}(t)] 
                               + \rho[(n_i+1)\beta_{n_i+1,i}(t) - n_i\beta_{n_i,i}(t)] 
			       + h_i\alpha_{n_i,i}(t) - f_i\beta_{n_i,i}(t), \nonumber \\
\end{eqnarray}
where the mRNA degradation rate is indicated by $\rho$. Notice that there is no interaction between different nuclei of the lattice. That is because we have neglected diffusion or transport effects here. Hence, one may consider each site as an independent unit and probability conservation holds for an individual site, such that
\begin{equation}
\sum_{n_i=0}^{\infty}\left[\alpha_{n_i,i}(t) + \beta_{n_i,i}(t)\right] = 1,
\end{equation}
at each site of the lattice. The introduction of the above property turns possible the use of exact solutions already available on the literature \cite{Innocentini2007}. Furthermore, we shall also use the symmetry properties underlying analyticity both as a mathematical tool and to give a biological interpretation for the model and its parameters \cite{Ramos2010}.

\subsection{Activity level, expected value and standard deviation}

In this manuscript we shall focus on the probability of finding the gene promoter site of the $i$-th nucleus at the active (repressed) state, indicated by $A_i(t)$ ($B_i(t)$), the mRNA average number, $\left\langle n_i \right\rangle(t)$, and, the standard deviation on $n_i$ and its consequence on positional precision of the borders of the stripe 2. The probabilities $A_i(t)$ and $B_i(t)$ are defined as
\begin{equation}\label{defprobs}
A_i(t) = \sum_{n_i=0}^{\infty}\alpha_{n_i,i}(t) \hspace{0.5cm} \text{and} \hspace{0.5cm} B_i(t) = \sum_{n_i=0}^{\infty}\beta_{n_i,i}(t).
\end{equation}
whereas the mRNA average number and the transcripts number fluctuations are, respectively, given by
\begin{equation}\label{defmeanvalue}
\left\langle n_i \right\rangle(t) = \sum_{n_i=0}^{\infty}{n_i}\left[\alpha_{n_i,i}(t) + \beta_{n_i,i}(t)\right]
\end{equation}
and
\begin{equation}\label{deffluct}
\sigma^2_i(t) = \left[\sum_{n_i=0}^{\infty}{n^2_i}\left[\alpha_{n_i,i}(t) + \beta_{n_i,i}(t)\right]\right] - \left[\left\langle n_i \right\rangle(t)\right]^2.
\end{equation}

Explicit expressions for $A_i(t)$, $B_i(t)$ and $\left\langle n_i \right\rangle(t)$ can be written in terms of the steady state probabilities to find the promoter site at the active and repressed states, respectively,
\begin{equation}\label{asymprobs}
a_i = \frac{f_i}{h_i + f_i} \hspace{0.5cm} \text{and} \hspace{0.5cm} b_i = \frac{h_i}{h_i + f_i}.
\end{equation}

The system approaches the steady state exponentially with a decay rate dependent on the constant
\begin{equation}\label{eps}
\epsilon_i = \epsilon = \frac{h_i + f_i}{\rho}.
\end{equation}

Note that we assumed this constant to have a fixed value along the AP-axis. This occurs because the quantity $\epsilon_i$ is an invariant of the model, as it was shown previously \cite{Ramos2010}. Then, the probabilities are written as
\begin{eqnarray}\label{probs}
A_i(t) &=& a_i +\left(A_i(0) - a_i\right)\text{e}^{-\epsilon \rho t}, \\
B_i(t) &=& b_i +\left(B_i(0) - b_i\right)\text{e}^{-\epsilon \rho t}.
\end{eqnarray}
where $A_i(0)$ and $B_i(0)$ are the initial conditions.

To show the time-dependent protein average numbers we define the constant $N = k/\rho$ in terms of which the average protein number at the steady state ($\mu_i$) is given, namely,
\begin{equation}\label{asymmeanvalue}
\mu_i = N\left(1-\chi\right)a_i + N\chi.
\end{equation}
and we get:
\begin{equation}\label{meanvalue}
\left\langle n_i \right\rangle(t) = \mu_i + U_i \text{e}^{-\rho t} + V_i \text{e}^{-\epsilon \rho t},
\end{equation}
where
\[
V_i = \frac{N(1-\chi)(A_i(0) - a_i)}{1-\epsilon}, \hspace{0.5cm} \text{and} \hspace{0.5cm} U_i = \left\langle n_i \right\rangle(0) - \mu_i - V_i.
\]
$\left\langle n_i \right\rangle(0)$ is the initial condition on the average mRNA numbers. Moreover, the fluctuation $\sigma^2(t)$ is given by
\begin{equation}\label{fluctfano}
\frac{\sigma^2_i(t)}{\left\langle n_i\right\rangle(t)} =  1 + \Delta_i(t)
\end{equation}
where $\Delta_i(t)$ is a position- and time-dependent term, {
namely
\begin{equation}\label{deltait}
\Delta_i(t) = \frac{\xi_i + \eta_i\text{e}^{-2\rho t} + \lambda_i\text{e}^{-\epsilon\rho t} + \psi_i\text{e}^{-(1+\epsilon)\rho t} + \phi_i\text{e}^{-2\epsilon\rho t}}{\left\langle n_i\right\rangle(t)}.
\end{equation}
The coefficients of the exponentials are written on the appendix.
}

The parameters $k$, $\chi$ and $\rho$ are assumed to be invariant along the AP-axis. We consider the mRNA synthesis rates to be mainly determined by the DNA sequence of the promoter region of the gene. For simplicity, we assume that the mRNA degradation has small interference of environmental conditions. That effective approach implies treating the RNA Polymerases or enzymes catalyzing mRNA degradation as uniformly distributed along the embryo.

The regulatory effect of the surrounding conditions around the $i$-th nucleus is introduced in terms of the gene switching rates, $h_i$ and $f_i$. If the concentrations of the regulatory proteins in a nucleus gene transcription the value of $f_i$ will be greater than $h_i$, and vice-versa. We may assume $h_i$ and $f_i$ to be dependent on space because such rates represent the effect of transcription factors (TF's). Furthermore, in literature we have not found any trace of auto regulation in the \textit{eve} mRNA stripe 2 formation during cycle 14A. One could also consider these rates to be time-dependent and, depending on the form of that dependence, the Eqs.\eqref{probs} and \eqref{meanvalue}) might not hold as solutions for the model.

\subsection{A model for the {\em eve}'s stripe 2 formation}

The Fig. \ref{Fig1} shows the comparison between the experimental data for estimated mRNA numbers (red lines) and the model predictions for the mRNA average numbers (green lines). The mRNA numbers are indicated at the vertical axis and the horizontal axis gives the percentage of the EL at AP direction. The red lines of the graphs are showing the mRNA average numbers at 9 different time instants for which experimental data is available.

The graphs at Figure 1 were obtained considering that the transcription dynamics consists of two phases. Each phase of the gene dynamics is characterized by a pair of switching rates, $f_i$ and $h_i$. The time-dependence of the switching rates can be represented in terms of Heaviside\footnote{The Heaviside function, $\Theta$, assumes two values, 0 or 1 if its argument is, respectively, negative or positive.} functions, $\Theta$, namely:
\begin{eqnarray}
\label{hef}
f_i(t) &=& \overline{f}_i\Theta(\tau_i - t) + \widetilde{f}_i\Theta(t - \tau_i)\\\nonumber
h_i(t) &=& \overline{h}_i\Theta(\tau_i - t) + \widetilde{h}_i\Theta(t - \tau_i),
\end{eqnarray}
where $\overline{h}_i$, $\overline{f}_i$, $\widetilde{h}_i$ and $\widetilde{f}_i$ are the switching constants at the $i$-th AP position (Figures S1 and S2 on Appendix). The parameter $\tau_i$, denominated {\em critical time}, is the time instant of the change of switching rates. This quantity may be space-dependent since it is associated with environmental regulatory factors which, in turn, are not distributed homogeneously along the AP axis. The parameters $k$, $\chi$ and $\rho$ will be the same during the two time phases.

We consider that $\epsilon = \frac{\overline{h}_i + \overline{f}_i}{\rho} = \frac{\widetilde{h}_i + \widetilde{f}_i}{\rho}$ is also invariant during time. Then, there will be only one free switching rate to be optimized. The assumptions above have a clear biochemical interpretation. Furthermore, they turn possible the exact calculation of the probability for the gene to be active and the mRNA average number, its standard deviation and the associated probability distributions.

The dynamics presented at Fig. \ref{Fig1} was obtained with the following equations for the dynamics of the probability for the gene to be active
\begin{equation}\label{ABit}
A_i(t) = \begin{cases}
                  \overline{a}_i + ( A_i(0) - \overline{a}_i)\text{e}^{-\epsilon \rho t}              &, 0 \leq t \leq \tau_i \\
                  \widetilde{a}_i + ( A_i(T_i) - \widetilde{a}_i)\text{e}^{-\epsilon \rho(t - \tau_i)}   &, t > \tau_i
         \end{cases}
\end{equation}
{
where the steady-state probabilities for the gene to be active are given as
\begin{equation}\label{asymprobsphase}
\underbrace{\overline{a}_i = \frac{\overline{f}_i}{\overline{h}_i +\overline{ f}_i}}_{\text{First phase}}, \hspace{0.5cm} \text{and} \hspace{0.5cm} \underbrace{\widetilde{a}_i = \frac{\widetilde{f}_i}{\widetilde{h}_i + \widetilde{f}_i}}_{\text{Second phase}}.
\end{equation}
The average number of mRNA's is given by
\begin{equation}\label{Mit}
\left\langle n_i \right\rangle(t) = \begin{cases}
				  \overline{\mu}_i + \overline{U}_i\text{e}^{-\rho t} + \overline{V}_i\text{e}^{-\epsilon \rho t}                &, 0 \leq t \leq \tau_i \\
					\widetilde{\mu}_i + \widetilde{U}_i\text{e}^{-\rho(t - \tau_i)} + \widetilde{V}_i\text{e}^{-\epsilon \rho(t - \tau_i)} &, t > \tau_i
		     \end{cases},
\end{equation}
with coefficients set as:
\begin{eqnarray*}
\overline{\mu}_i = N\left(1-\chi\right)\overline{a}_i + N\chi, \hspace{0.5cm} \widetilde{\mu}_i = N\left(1-\chi\right)\widetilde{a}_i + N\chi, 
\end{eqnarray*}
\begin{eqnarray*}
\overline{V}_i = \frac{N(1-\chi)(A_i(0) - \overline{a}_i)}{1-\epsilon}, \hspace{0.5cm} \widetilde{V}_i = \frac{N(1-\chi)(A_i(\tau_i) - \widetilde{a}_i)}{1-\epsilon},
\end{eqnarray*}
\begin{eqnarray*}
\overline{U}_i = \left\langle n_i \right\rangle(0) - \overline{\mu}_i - \overline{V}_i, \hspace{0.5cm} \widetilde{U}_i = \left\langle n_i \right\rangle(\tau_i) - \widetilde{\mu}_i - \widetilde{V}_i,
\end{eqnarray*}
where $A_i(0)$ and $\left\langle n_i \right\rangle(0)$ are the initial conditions of the dynamics and $A_i(\tau_i)$ and $\left\langle n_i \right\rangle(\tau_i)$ are the initial conditions for the second phase.
}

\subsection{Initial conditions}

The initial conditions of the average mRNA numbers for the first phase are built by fitting of experimental data at the time class C13 with the function:
\begin{equation}\label{Mi0}
\left\langle n_i \right\rangle(0) = c_0 + c_1\text{e}^{-c_2(i-35.5)}
\end{equation}
where $c_0 = 1.36468$, $c_1 = 12.3591$ e $c_2 = 0.0575796$. The use of the exponential function replaces dozens of data for conditions at instant C13 and represents a great reduction on the number of parameters of the model.

{
The initial condition on $\left\langle n_i \right\rangle(0)$ given by the Eq. (\ref{Mi0}) was obtained in terms of $\left\langle n_{i=41.5} \right\rangle(0)$. We consider the experimental value in 41.5\% AP position at C13 as initial condition for the number of mRNA, that is, $\left\langle n_{i=41.5} \right\rangle(0)=10.6892$. As for the initial condition $A_{i=41.5}(0)$ at 41.5\% AP position, we are far from any estimate because of scarcity of information. Nevertheless, since our approach is based on a Markov process, the tendency is that, during the time evolution, the system stops feeling the effect of the initial condition and, because of this, the absence of an estimate for $A_{i=41.5}(0)$ — or even for $\left\langle n_{i=41.5} \right\rangle(0)$ — does not represent a problem.
}

An alternative to estimate $A_i(0)$ is to consider that the system is described approximately by an arbitrary steady state of the model. In that case the initial conditions for the probabilities for the gene to be active is obtained from $\left\langle n_i \right\rangle(0)$. That results into
\begin{equation}\label{Ai0}
A_i(0) = \frac{\left\langle n_i \right\rangle(0)-N\chi}{N(1-\chi)}.
\end{equation}
Notice that the space dependence of $A_i(0)$ is inherited from $\left\langle n_i \right\rangle(0)$.

\subsection{Parameters estimation}

The parameters of the model have been obtained by optimization. The values of $k$, $\chi$ and $\rho$ were obtained by fitting the dynamics at the position 41.5\% of the EL. Then, we set those values along the whole AP axis because they are indicating characteristics that are intrinsic to the promoter site and the mRNA molecules.

{
  
  The parameter $\chi$ quantifies the proportion between the two mRNA production modes. It may be estimated as 0 simply because the phenomenon is being described in terms of two well characterized activation levels: one with high mRNA production rate and another with low production rate (which, in first approximation, may be considered null).

  The estimates for the probabilities $\overline{a}_{i=41.5}$ and $\widetilde{a}_{i=41.5}$ may also be conceived with the help of experimental data and the “all-or-nothing” character of our approach. At 41.5\% AP position the production of \textit{eve} mRNA is considerably greater than at other AP positions. Furthermore, since the \textit{eve} mRNA stripe 2 formation occurs in wild-type embryos, the tendency of biological system is always to produce \textit{eve} mRNA during the first temporal classes. This means that, in first approximation, we may estimate $\overline{a}_{i=41.5} \sim 1$. On the other hand, the biological system starts ceasing the production of \textit{eve} mRNA after $\tau_i$. Hence, we may estimate $\widetilde{a}_{i=41.5}=0$ as the probability for the gene being active at the end of the second phase.

  By estimating $\chi = 0$, $\overline{a}_{i=41.5} = 1$, $\widetilde{a}_{i=41.5}=0$ and $\left\langle n_{i=41.5} \right\rangle(0)=10.6892$, optimization preliminary tests in 41.5\% AP position provide $k = 0.0701998${sec}$^{-1}$, $\epsilon = 1.43301${sec}$^{-1}$, $\rho = 0.00242665${sec}$^{-1}$, $\tau_{i=41.5} = 2053.09${sec} and $A_{i=41.5}(0) = 0.0165307$. By inserting these optimized values, the initial condition $\left\langle n_{i=41.5} \right\rangle(0)$ and the estimates for $\chi$, $\overline{a}_{i=41.5}$, $\widetilde{a}_{i=41.5}$ into Eqs. (\ref{ABit}) and (\ref{Mit}), the average number $\left\langle n_{i=41.5} \right\rangle(t)$ may be obtained and compared to the experimental data, as Figure \ref{Fig2} shows. 
}

The remaining parameters have been computed by analysis of the experimental data along the space and it results:
\begin{eqnarray}
\label{ansatzespacial}
\overline{a}_i   &=& \text{exp}\left[-0.04(i - 41)^2\right]\\
\widetilde{a}_i  &=& 0                                     \\
\tau_i           &=& \tau_{41.5}\text{exp}\left[-0.04(i-41)^2\right] \label{critical}
\end{eqnarray}
where $\tau_{41.5} = 2053.09$ sec.

\begin{figure}
\centering
\includegraphics[width=0.5\linewidth]{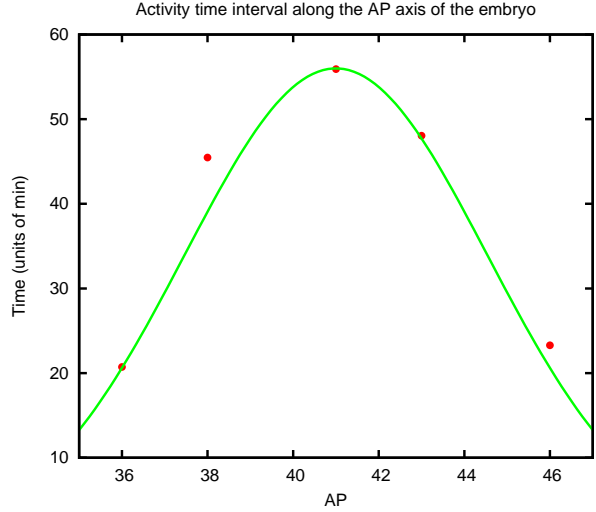}
\caption{A comparison between the Gaussian shape for critical time as obtained with the Eq. (\ref{critical}) and experimental data on transcriptional activity around the peak of expression of {\em eve}'s stripe 2, using experimental data presented in Ref. \cite{Bothma14}. The experimental data are shown as the solid red circles and the green lines are indicating the model's results.}
\label{figactv}
\end{figure}

The Gaussian profile of $\overline{a}_i$ is based on fact of that, at sufficient long time, the spatial distribution of the number of transcripts will be very similar to the space-dependence of the asymptotic activation probability (see Eqs.\eqref{probs} and \eqref{ansatzespacial}). The Gaussian profile of $\tau_i$ is related to the sharpening of the stripe 2 as follows: The TF's that repress the transcription of the {\em eve} gene are getting produced on both sides of the stripe 2; Their concentrations achieve sufficiently large values at the critical time instants given by the Eq. (\ref{critical}); As a consequence, the probability for the gene being repressed increases in accordance with the increase on the amount of repressive TF's. Indeed, the reduction of the transcriptional activity of the {\em eve}'s enhancer controlling the stripe 2 formation has a Gaussian profile, as it is shown on Fig. \ref{figactv}. 

The amplitudes of the functions $\overline{a}_i$ and $\tau_i$, and the estimate of $\widetilde{a}_i$, are obtained by optimization. The centers of both Gaussian functions $\overline{a}_i$ and $\tau_i$ are at 41\% of EL because the expression levels reached at both 40.5\% of EL and 41.5\% of EL are very close to one another so that we assume that the center of the stripe is at the midpoint between these AP positions.

\section{The fluctuations of mRNA number}

\begin{figure}
 \centering
 \includegraphics[width=0.5\linewidth]{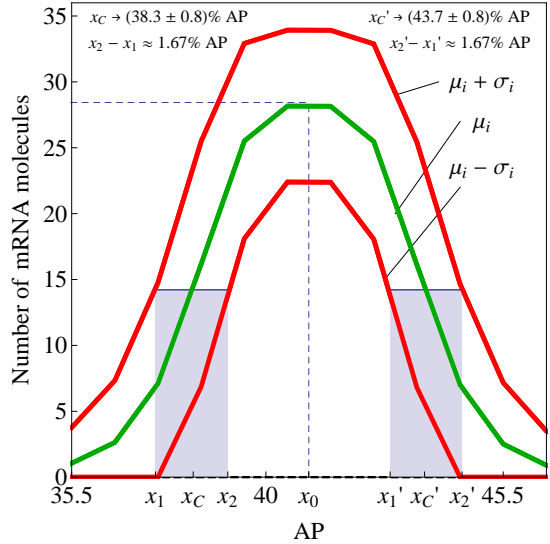}
 \caption{The implications of the fluctuations on the mRNA numbers on the position of the anterior and posterior borders of {\em eve}'s stripe 2 domain at time class T5. We consider three particular curves, one (green color) for the average number of mRNA synthesized by the {\em eve}'s gene and the remaining two (red color) given by the average plus or minus the standard deviation on the number of transcripts.} 
\label{spacefluct} 
\end{figure}

{
The Eq. \eqref{fluctfano} gives the general form of the noise obtained with the stochastic binary model. Since we have two phases in our model, the coefficients for the exponentials on the definition of $\Delta_i(t)$ at each time phase are defined as follows. For $0\leq t \leq \tau_i$, the terms $\xi_i$, $\eta_i$, $\lambda_i$, $\psi_i$ and $\phi_i$ become, respectively, $\overline{\xi}_i$, $\overline{\eta}_i$, $\overline{\lambda}_i$, $\overline{\psi}_i$ and $\overline{\phi}_i$, given by
\begin{eqnarray*}
\overline{\xi}_i     &=& \frac{N^{2}(1-\chi)^{2}\overline{a}_i(1 - \overline{a}_i)}{1 + \epsilon} \\\nonumber
\overline{\eta}_i    &=& \left[{\left\langle n_i\right\rangle_0}^{2} - {\overline{\mu}_i}^{2} \right] + \left[\left\langle n_i\right\rangle_0 - \overline{\mu}_i \right] + \frac{N^{2}(1-\chi)^{2}}{\epsilon - 1}\left[A_i(0)(1 - A_i(0)) - \overline{a}_i(1 - \overline{a}_i)\right]\\\nonumber
                     &+& \left\{\frac{2N(1+\chi)}{\epsilon - 2} + 2\left[N(1 - \overline{a}_i + \chi\overline{a}_i) - \left\langle n_i\right\rangle_0\right]-1\right\}\overline{V}_i - (2\overline{\mu}_i +1)\overline{U}_i - {\overline{U}_i}^2 \\\nonumber
\overline{\lambda}_i &=& \frac{2\overline{V}_i}{2 - \epsilon}\left[ N(1 + \chi - \epsilon) + \overline{\mu}_i(\epsilon - 2) + \overline{a}_i\epsilon(1 - \chi) \right] \\\nonumber
\overline{\psi}_i    &=& \frac{2N^{2}(1-\chi)}{1 - \epsilon^{2}}\left[ A_i(0)(1 - A_i(0)) - \overline{a}_i(1 - \overline{a}_i) \right] + 2 \left[ \overline{\mu}_i - N(1 - \overline{a}_i + \chi\overline{a}_i) + \overline{V}_i \right]\overline{V}_i  \\\nonumber
\overline{\phi}_i    &=& - \overline{V}_i.
\end{eqnarray*}
For $\tau_i<t$, the terms $\xi_i$, $\eta_i$, $\lambda_i$, $\psi_i$ and $\phi_i$ become, respectively, $\widetilde{\xi}_i$, $\widetilde{\eta}_i$, $\widetilde{\lambda}_i$, $\widetilde{\psi}_i$ and $\widetilde{\phi}_i$, which correspond to the asymptotic probability $\widetilde{a}_i$ and ``initial conditions'' $A_i(\tau_i)$ and $\left\langle n_i \right\rangle(\tau_i)$ instead of $\overline{a}_i$, $A_i(0)$ and $\left\langle n_i \right\rangle(0)$. Moreover, these terms are evaluated at $t - \tau_i$ instead of $t$. That is analogous to the procedure proposed for the evaluation of the mRNA average values.
}

The Eq. \eqref{fluctfano} shows the prediction of the model for the dynamics of the ratio of the standard deviation for the mRNA mean number, also called the Fano factor. The Fano factor is useful to determine how different from a Poissonian a probability distribution is: a Poisson distribution has a Fano factor equals to one. A super Fano (or sub Fano) probability distribution has Fano factor greater (or smaller) than one. Since the quantity $\Delta_i(t)$ obeys $\Delta_i(t) \ge 0$ for all $i$ and $t$, it causes the Fano factor to be equals to or greater than one. Therefore, the enhancer controlling formation of the eve's stripe 2 along the AP axis during the cycles 13 and 14 leads the {\em eve} gene to behave as a Fano or super Fano source of mRNA's. 
  
The standard deviation on the mRNA numbers during cycles 13 and 14 along the AP axis is given by the vertical bars at the plots of the Figure \ref{Fig1}. Accordingly to the Eq. \eqref{fluctfano}, the standard deviation is given by $\sigma_i(t) = \sqrt{\left\langle n_i \right\rangle(t)} \sqrt{ 1 + \Delta_i(t)}$. At the initial time instants the noise values are greater at positions where $\left\langle n_i \right\rangle(t)$ assumes its maximal values. When the stripe 2 starts to achieve its shape the noise has three different characteristics: it is almost zero on the regions at the right of the stripe; it has intermediary values around the peak of expression; the maximal values of the noise are exhibited at positions on the borders of the stripe. During the two final time classes the noise amplitude is  greater at the regions where $\left\langle n_i \right\rangle(t)$ is higher. The difference on the dynamic behavior of the noise is due to the values of the asymptotic values, $\Delta_i(\tau_i)$ and $\Delta_i(\infty)$, and the differences $\Delta_i(\tau_i)-\Delta(0)$ and $\Delta_i(\infty) - \Delta_i(\tau_i)$. That is because the system approaches the equilibrium with the same speed at all AP positions and phases of the dynamics.

The Fig. \ref{spacefluct} shows a plot for the average mRNA numbers and their fluctuations on the region from 35.5\% EL to 46\% EL at the time class T5. The green line indicates the values of $\langle n_i \rangle$ while the values of $\langle n_i \rangle \pm \sigma_i$ are indicated by the red lines. The dashed blue line indicates the position where the mRNA is maximal, $i=41$. There are two rectangles with top horizontal blue lines and grey colored area around the positions $x_C$ and $x_C'$, with a base length going from $x_1$ to $x_2$ and $x_1'$ to $x_2'$, respectively. The height of the two rectangles is half of the maximum number of mRNA's. The spatial fluctuation on the position of the {\em eve}'s stripe 2 can then be inferred from the fluctuations on the mRNA numbers, $\sigma_i$. The two rectangles are indicating the spatial variation on the formation of the anterior and posterior borders of the stripe 2. Based on our results, the position of the anterior border of the stripe 2 is at $x_C \pm 0.8\%$ while the posterior border of the stripe 2 is at $x_C' \pm 0.8\%$.

\section{Discussion}

\subsection{Parameters of the model and fitting}

An important characteristic of our approach is the small number of parameters. The fitting shown at Fig. \ref{Fig1} has required 14 parameters for 42 $\times$ 9 (= 378) experimental data. The good agreement between the model and experimental data is mainly due to fitting at intermediary and not at final instants of the process. That is because the initial and final states of each phase are already specified (by initial conditions and rates $h_i$ and $f_i$, respectively). Therefore, the fitting presented at Fig. \ref{Fig1} is an actual test for the dynamics resulting from our model.

{
The Fig. (\ref{Fig2}) shows not only that proposed approach is able to describe experimental with good agreement but also that it is possible to find a set of parameters (for $k$, $\chi$, $\rho$, $\epsilon$, $\overline{a}_i$, $\widetilde{a}_i$, $\tau_i$ and $A_i(0)$) coherent with the model and literature. Thanks to the model's good performance in describing the experimental data at 41.5\% AP position we have extended it to describe the numbers of {\em eve} dynamics to other AP positions.
}

\subsection{Estimates of the mRNA numbers and parameters optimization}

The analysis of the Fig. \ref{Fig1} may lead us to question about the correctness of the estimates of the \textit{eve} mRNA numbers. Indeed, we expect that a precise and direct detection of the \textit{eve} transcripts would require an adjustment of our estimate. However, it is fair to assume the preservation of the qualitative aspects of the mRNA number dynamics. In that case, the refinement of the model's parameters would be required for fitting data but biological interpretation of the model would remain useful to understand the role of the fluctuations during \textit{Drosophila} embryo development.

{
 In the literature, we have found that size of the \textit{eve} coding sequence is 1477 bp \cite{StPierre2014} and the rate at which the RNA-Polymerase II transcribes a sequence is 1400 bp/min = 23.33 bp/sec \cite{Shermoen1991,Thummel1992,Femino1998}. This allows to estimate the time of production of one \textit{eve} mRNA in 63.3 seconds, which corresponds to the rate $k = 1/(63.3 \text{ sec}) = 1.579\cdot 10^{-2} \text{ sec}^{-1}$. This estimate is also consistent with other values \cite{Bolouri2003}.
 
 The estimate of the degradation rate $\rho$ of the \textit{eve} mRNA is based on the half-live of the \textit{ftz} mRNA. This is adopted due to scarcity of data about \textit{eve} mRNA. In literature \cite{Edgar1986,Surdej1994,Riedl1996}, the half-live of \textit{ftz} mRNA is between 6 and 7 minutes, which corresponds to a degradation rate of order of $10^{-3} \text{ sec}^{-1}$. This value establishes the order of magnitude for the degradation rate $\rho$ of the \textit{eve} mRNA.
}

\subsection{The two phases dynamics}

The graph at Fig. \ref{Fig2} suggests that the transcription of the \textit{eve} gene around the position 41.5\% EL is a two-phases process. As one may verify at the Eq.\eqref{Mit} the two phases have different attractors, the first phase has $\overline{\mu}_i$ as a fixed point while $\widetilde{\mu}_i$ is the fixed point of the second phase. The average number of mRNA's has direct dependence on the asymptotic probabilities, $a_i$ and $b_i$, as it is shown at the Eq. \eqref{asymmeanvalue}. Furthermore, $\mu$ also depends on $k$, $\chi$, and $\rho$ which are fixed constants. Therefore, the occurrence of two attractors is due to the existence of two asymptotic probabilities, each of them associated with one temporal phase of the dynamics. The asymptotic probabilities are functions of the switching parameters, as shown at the Eq. \eqref{asymprobsphase}. The regulatory configuration of the promoter site, despite its random switching, is characterized by a pair $f_i$ and $h_i$ that leads it to be mostly active until $\tau_i$. After it, the promoter site shall be mostly repressed and mRNA degradation will be the prevailing reaction.

The introduction of the Heaviside time-dependence on the switching rates and the space dependence is a generalization to previous versions of this model. That permits introduction of diffusion or other space dependent processes. The use of Heaviside functions for the switching rates turns possible to decompose the temporal domain in an arbitrary number of slices and the solution is an analogous to a Riemann sum. Then, the exact solution presented here is a generalization of existing solutions to this model. 

The two-phase behavior of the \textit{eve}'s promoter site is caused by the interaction of multiple TF's with the enhancer controlling the stripe 2 formation. The interaction of the TF's along the AP axis is orchestrated in such a way that the \textit{eve} gene promoter site operation is effectively binary. That behavior of the \textit{eve} promoter supports the proposition of the model of Eq. \eqref{mastereq} as a tool to understand the role of random fluctuations during \textit{Drosophila} development. It provides the simplest approach to regulated gene transcription, is fully solvable, and the symmetries underlying its analyticity have been characterized. Here, the existence of symmetries has a practical implication as it allows us to take $\epsilon$ as an invariant of the model on both space and time. That assumption leads to a reduction of the free parameters of the model and contributes to avoid the over-fitting.

\subsection{Phenomenology of the model}

The invariance of $\epsilon$ can be interpreted biochemically as the characteristic time of the promoter's site capability for the transition between the active and repressed states in relation to the characteristic time of the mRNA degradation. For high (low) values of $\epsilon$, a transition cycle will be completed fastly (slowly) when compared to mRNA degradation.  Here, $f_i$ and $h_i$ relates to the time fraction of a switching cycle that the promoter site will take to switch to the active and repressed states, respectively. For $f_i > h_i$, the promoter will be predominantly at the active state while the repressed state is most likely for the opposite condition.

The invariance of $\epsilon$ also helps on the understanding of the enhancer-TF's interaction. For $f_i >> h_i$, the promoter site is highly active, and the majority of the enhancer-TF's interactions contributes to stimulate \textit{eve}'s transcription. Hence, one may suppose the enhancer inducing an active state of the promoter during the first phase of the stripe 2 formation. The initial phase has higher concentrations of Bcd and Hb that contribute for such a higher activation. As the gap genes are expressed, repressive TF's (Gt, Kr and Kni) start binding to the enhancer and the balance between $f_i$ and $h_i$ changes towards $h_i >> f_i$. Then, the promoter site is expected to be mostly repressed.

The critical time can be interpreted as a time instant when there is a saturation of repressive TF's bound to the enhancer and the promoter is most likely to be repressed. There is a high number of possible configurations for the stripe 2 enhancer. Hence, we expect a smooth transition from the initial configuration to the asymptotic values. This is represented by the exponential time decays of average mRNA number and the promoter state.

The Fig. \ref{figactv} shows a comparison between the Gaussian profile as proposed for the critical times and the observed data for the transcriptional activity time interval of the {\em eve} gene as detected experimentally at some positions around the peak of expression \cite{Bothma14}. It is remarkable that the Gaussian curve is in good agreement with the analyzed data. The transcriptional activity is directly related to the binding of repressive TF's to the enhancer controlling the formation of the stripe 2 of the {\em eve} gene. The increase on the amount of repressive TF's reduces the transcriptional activity by their binding to the enhancer. Furthermore, the spatial form of the activity reduction follows a Gaussian spatial pattern centered around the peak of expression, as it happens with the critical times that are given by the Eq. \eqref{critical}.

\subsection{The intrinsic noise on mRNA numbers due to promoter dynamics}

Our model can also be used as a tool to understand the characteristics of the noise on the number of mRNA's when it is produced by a two levels system. The form of the noise presented at the Eq. \eqref{fluctfano} is similar to the form of the translational noise as introduced in a paper considering a two stages model to stochastic gene expression \cite{Thattai2001}. The noise on the protein numbers at the steady state regime was written as $\sigma^2 / \left \langle n \right \rangle = 1 + B$, where $B$ is a constant denoted as the bursting size \cite{Shahrezaei08}. One may show that our two levels model also has a bursting limit, with burst size $\Delta(t)$. The bursting occurs at the limit of many very fast mRNA synthesis during the very brief time interval of duration of the ``on'' state of the gene. Even though $B$ and $\Delta(t)$ are called, respectively, translational and transcriptional burst sizes, they are actually average values. That is, either the burst amplitude and its time duration vary randomly at each burst and, hence, the burst size has a different value at each bursting event.

The above discussion on noise is important to understand recent results of a single molecule detection experiment that was performed to evaluate the dynamics of the formation of the stripe 2 of the {\em fruit fly} embryos \cite{Bothma14}. The amount of eve's mRNA molecules synthesized per transcriptional event was also probed. It was shown that the {\em eve} molecules are synthesized in a burst-like regime with burst size and duration varying randomly at each transcriptional event. Based on that observation, the authors have concluded that a two level model to gene expression would not be capable of accounting this phenomena and, therefore, the transcription of the eve gene should occur at multiple levels of synthesis rate. The reasoning underlying that conclusion is that each level of gene transcription has a single burst size corresponding to it. That conclusion would be well supported by the assumption of {\em deterministic} binding of the RNAP to the DNA. Then, for each burst size there will be a given amount of RNAP bound to the DNA. However, the number of reacting molecules inside the cell is small and it leads the intracellular phenomena to be stochastic \cite{Delbruck1940}. Hence, the observation of the multiples burst sizes is not necessarily due to multiple levels of the transcription of the {\em eve} gene but a manifestation of the inherent stochasticity of biological phenomena. Particularly, the randomness of the {\em eve}'s stripe 2 formation. Furthermore, since the burst size in our two levels model is both random and time-dependent, as aforementioned, we consider that it is very early to discard it as a good model to evaluate noise in {\em fruit fly} developmental processes.

\subsection{Variation on the stripe 2 borders position}

The analysis of the noise on the mRNA numbers has turned possible for us to investigate its implications on the positional precision of the control of: The location of the anterior and posterior borders of the stripe 2; The fluctuations on the domain width. Our results indicate that the noise on the position of the borders of the domain is of 0.8\% of the EL which is in good agreement with experimental results \cite{Surkova2008a}. In that reference, the noise on the position of the {\em eve} domain is investigated for time classes until T5 and it is shown to be of the order of 1\% of the EL. We have also calculated the variation on the domain size and it goes from 3.8\% EL to 7\% EL, and its average width is 5.4\% EL.

In this manuscript, the noise on the position of the border of the domain of the stripe 2 have been evaluated in terms of a model for the fluctuations on the number of {\em eve} molecules. Despite it may appear as a limitation of our approach it has a useful consequence, as it suggests the possible inferior values of the positional error of the segmentation process. In our approach, the noise on the position of the peak of expression of the {\em eve} stripe 2 is not considered and it would be interesting to incorporate it here. Similarly, one may also investigate the role of the fluctuations on the numbers of the TF's regulating the expression of the stripe 2. Furthermore, the usual error of the experimental data is not included in our model. Therefore, it is acceptable that our model's predictions for the positional noise on the stripe 2 borders is smaller than that observed experimentally ($\sim 0.8\%$ EL).

{
  There is an interesting implication for the good agreement between the predicted spatial variation of the borders position due to the intrinsic noise arising from the promoter and the existing empirical data \cite{Surkova2008a}. The intrinsic noise due to the promoter's dynamics might account for all of the observed spatial variability, and other extrinsic noise sources are being damped or filtered by mechanisms that are to be determined \cite{Gregor2007,Ramos2015}. One must notice, however, that our predictions for the noise are related with our extrapolation of mRNA numbers that are not from the {\em eve} gene. It is natural to expect that the mean number of mRNA's may differ from gene to gene. Naturally, that would cause a change on the values of the parameters of the model and the noise predictions. Nevertheless, the theoretical machinery does not assume anything about the precise numbers of even-skipped molecules and so the model will still be valid whenever such measurements become available.
}

\begin{acknowledgments}
The authors are thankful to Prof. John Reinitz for invaluable discussions. We thank the two anonymous reviewers for suggestions that increased the quality of our manuscript. GNP was supported by PNPD-CAPES under the Complex Systems Modeling Grad School (EACH – USP). The authors are thankful to FAPESP and CAPES for financial support and to Institute of Physics of S\~ao Carlos (IFSC – USP).
\end{acknowledgments}

\appendix

\section{}

{
\par\textbf{Digitalization}
\\

The experimental data were digitalized by using the program \textit{DigiteIt 1.5}. 
\\

\par\textbf{Optimization}
\\

All the fits obtained by optimization ran via least-squares method by using the Wolfram Mathematica 7.0 program.
\\
\par \textbf{\textit{Ansatz} of a two-phase transcriptional dynamics}
\\

\textit{Heaviside function}. For those unfamiliar with the Heaviside function, we present at the Figure \ref{FigS1} the time dependence of the switching parameters $f_{i}(t)$ and $h_{i}(t)$ as presented at the Eq.(10) of our manuscript. Initially, the pair $(\overline{f}_{i}(t), \overline{h}_{i}(t))$ dominates the transcriptional dynamics until a critical instant $T_i$. From that time instant the dynamics is guided by the pair of rates $(\widetilde{f}_{i}(t), \widetilde{h}_{i}(t))$.
\\
\begin{figure*}
 \centering
 \includegraphics[scale=0.8]{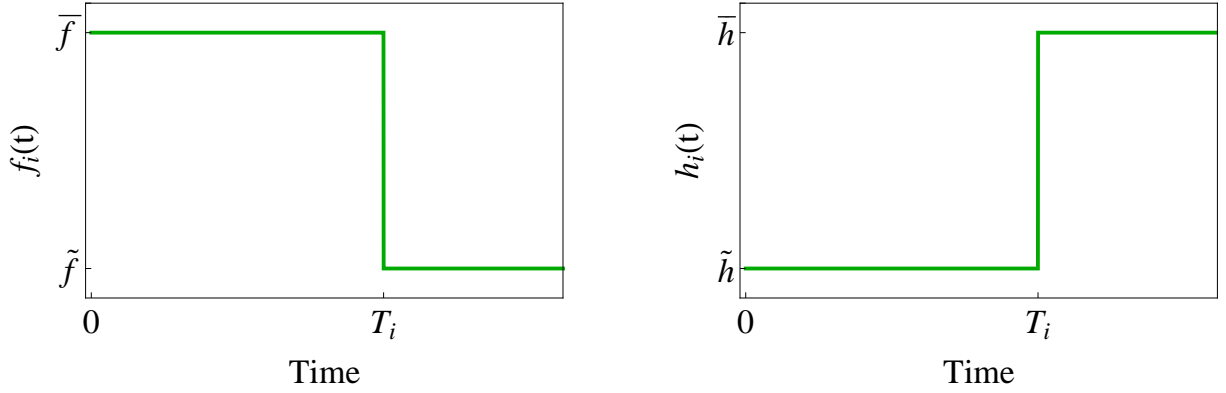}
 \caption{\label{FigS1} 
   Heaviside-like time-dependence in the rates $h_i$ and $f_i$. Each phase of the dynamics of the mRNA expression at 41.5\% AP position is associated with an evolution toward seemingly asymptotic level. This, in turn, is associated with a pair of transition rates: the regulatory effect is represented by $\overline{h}_i$ and $\overline{f}_i$ in the first phase and by $\widetilde{h}_i$ and $\widetilde{f}_i$ in the second phase. Since the first and second phases are respectively characterized by the formation and disappearance of the stripe 2, this means that whereas the gene performs the transition \textit{off} $\rightarrow$ \textit{on} faster than \textit{on} $\rightarrow$ \textit{off} during the first phase, the inverse occurs during the second phase. Mathematically, this means that $\overline{f}_i >\widetilde{f}_i$ and $\widetilde{h}_i < \overline{h}_i$.} 
\end{figure*}

\textit{Parameters optimization}. For optimization of the switching parameters purposes  one may work with asymptotic probabilities for the gene to be active and also represent these probabilities in terms of a Heaviside function. We start showing the asymptotic probabilities for the gene to be active at each time phase as
\begin{matriz}\label{asymproba}
\overline{a}_i = \frac{\overline{f}_i}{\epsilon \rho} \hspace{0.5cm} \widetilde{a}_i = \frac{\widetilde{f}_i}{\epsilon \rho}.
\end{matriz}
Note that we are using the constants $\epsilon$ and $\rho$ instead of $\overline{h}_i + \overline{f}_i$ or $\widetilde{h}_i + \widetilde{f}_i$. That turns the optimization process simpler.

The optimization process is greatly simplified by the introduction of the Eq.\eqref{asymproba}, which can be written as:
\begin{matriz}\label{qit}
q_i(t) = \begin{cases}
                  \overline{a}_i   &, 0 \leq t \leq T_i \\
                  \widetilde{a}_i  &, t > T_i
         \end{cases}
\end{matriz}
and is shown at the Figure \ref{FigS2}. The simplification is due to the fact that $q_i(t)$ only assumes real values between 0 and 1 while the switching rates may assume an non-negative real value.
\begin{figure*}
 \centering
 \includegraphics[scale=0.8]{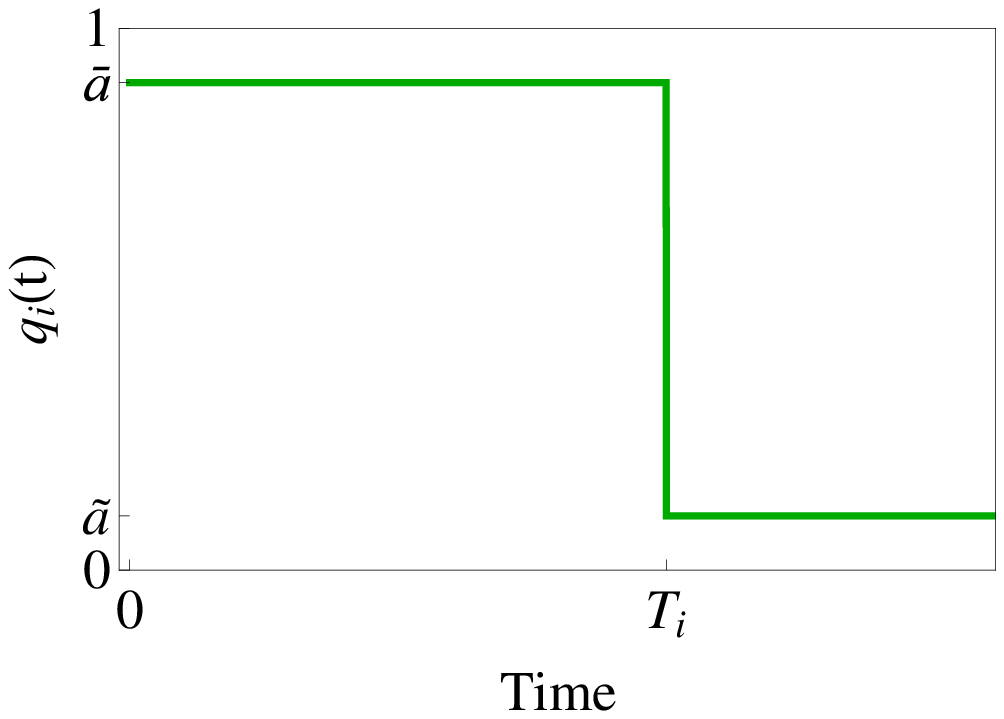}
 \caption{\label{FigS2} {
     Heaviside-like time-dependence in the rates $q_i$. Heaviside-like time-dependence in rates $h$ and $f$ also makes the asymptotic probabilities $\overline{a}_i$ and $\widetilde{a}_i$ vary suddenly over time. It is convenient to work with the $q_i$, given by $q_i = (\epsilon \rho)^{-1}f_i$, because it has units of probability, that is, $0 \leq q_i \leq 1$, which is useful to investigate the regulatory information and to run optimizations for parameters of the model.}} 
\end{figure*}
\\
\par \textbf{Coefficients for the Eq. (\ref{deltait})}
\\

\begin{eqnarray*}
{\xi}_i     &=& \frac{N^{2}(1-\chi)^{2}{a}_i(1 - {a}_i)}{1 + \epsilon} \\\nonumber
{\eta}_i    &=& \left[{\left\langle n_i\right\rangle_0}^{2} - {{\mu}_i}^{2} \right] + \left[\left\langle n_i\right\rangle_0 - {\mu}_i \right] + \frac{N^{2}(1-\chi)^{2}}{\epsilon - 1}\left[A_i(0)(1 - A_i(0)) - {a}_i(1 - {a}_i)\right]\\\nonumber
                     &+& \left\{\frac{2N(1+\chi)}{\epsilon - 2} + 2\left[N(1 - {a}_i + \chi{a}_i) - \left\langle n_i\right\rangle_0\right]-1\right\}{V}_i - (2{\mu}_i +1){U}_i - {{U}_i}^2 \\\nonumber
{\lambda}_i &=& \frac{2{V}_i}{2 - \epsilon}\left[ N(1 + \chi - \epsilon) + {\mu}_i(\epsilon - 2) + {a}_i\epsilon(1 - \chi) \right] \\\nonumber
{\psi}_i    &=& \frac{2N^{2}(1-\chi)}{1 - \epsilon^{2}}\left[ A_i(0)(1 - A_i(0)) - {a}_i(1 - {a}_i) \right] + 2 \left[ {\mu}_i - N(1 - {a}_i + \chi{a}_i) + {V}_i \right]{V}_i  \\\nonumber
{\phi}_i    &=& - {V}_i.
\end{eqnarray*}
}



%
%
\end{document}